\documentclass[10pt,prl,aps,twocolumn,floatfix]{revtex4}
\usepackage{graphicx}
\def\be{\begin{equation}}
\def\ee{\end{equation}}
\def\ba{\begin{eqnarray}}
\def\ea{\end{eqnarray}}
\def\half{\frac{1}{2}}
\def\det{\mbox{det}}
\def\sf{\sqrt{f}}
\def\D{\Delta}
\def\cF{{\cal F}}
\def\iome{\mbox{Im~}\omega}
\begin{document}
\title{Critical current of a superconducting wire via gauge/gravity duality}
\author{Sergei Khlebnikov}
\affiliation{Department of Physics, Purdue University, West Lafayette, IN 47907}
\begin{abstract}
We describe application of the gauge/gravity duality to study of
thin superconducting wires at finite current. 
The large number $N$ of colors of the gauge theory is identified with the number of
filled transverse channels in the wire. On the gravity side, the physics is
described by a system of D3 and D5 branes intersecting over a line. We consider
the ground state of the system at fixed electric current and find that at zero 
temperature the normal
state is always unstable with respect to appearance of a superconducting component.
We discuss relation of our results to recent experiments on statistics of the switching 
current in nanowires.
\end{abstract}

\maketitle
Destruction of superconductivity by a current flowing through the superconductor has long
been recognized as an important topic both from the viewpoint of fundamental physics and
for applications (for a
review, see Bardeen \cite{Bardeen}). One expects that theoretical treatment should be 
the simplest for samples that are effectively one-dimensional (1d), i.e., wires in which
the superconducting density depends on only one coordinate. 
Even for this case, 
however, a complete theoretical treatment of the transition at {\em fixed current}
has not been forthcoming. The main difficulty
lies, somewhat ironically, in the definition of a current-carrying {\em normal} state:
the equilibrium Fermi distribution used by the conventional mean-field theory 
is clearly inadequate for the purpose as it carries
no current at all. 

A related but different problem is a transition caused not by a fixed current 
(sustained by an external battery) but by an {\em initial winding} of the order 
parameter.
Experimentally, this is the condition appropriate for a thin superconducting ring.
In this case, the supercurrent is only metastable: there are fluctuations that
lower the free energy \cite{Little}
(provided there is an amount of disorder or a finite 
temperature \cite{Khlebnikov}); these have become known as phase slips.
The relevant question to ask then
is whether there is a maximal winding number density beyond which
the metastability becomes classical instability.

To define a current-carrying normal state, one needs 
to include some mechanism that
equilibrates the electrons with respect to momenta.
One possible approach is to include weak 
scattering via a kinetic equation. Another, which we adopt here, is to start at the
opposite extreme---a theory with strong electron-electron interactions.
One may hope that for such a theory
there is a complementary (dual) description in terms of weakly 
coupled collective modes. If these collective modes are the same
as seen on the superconducting side, one will have a unified description applicable 
to both phases, which should make understanding the phase transition easier.

Recently, following the discovery of the gauge/gravity duality \cite{Maldacena:1997re},
carrying out this program has become practical. The gauge/gravity duality allows
one to study a strongly coupled $SU(N)$ gauge theory with a large number $N$
of colors by doing calculations in classical
gravity, albeit in a higher-dimensional spacetime. Here we use this method
to study a strongly-coupled thin superconductor at fixed current.
The number $N$ of colors is taken to correspond to the number of 
populated transverse modes (channels) in the wire. We find that, at zero
temperature, the normal-only state is always unstable with respect to appearance of 
a superconducting component. At sufficiently large currents, $J > J_c$, the instability
occurs only for modes with nonzero winding of the order parameter. We interpret this 
as an indication that at $J > J_c$ at least a part of the total current must be a 
supercurrent.

The absence of a depairing transition at large currents is surprising. It may 
have to do with our system being perfectly momentum-conserving. That, for instance, 
precludes the Landau process---production of quasiparticles with momenta antiparallel to
the flow. The situation may change at a finite temperature, due to the presence of
thermal quasiparticles. Indeed, as
we discuss towards the end, the
idea of a near-critical behavior at some finite current
is consistent with the results of
recent experiments \cite{Sahu&al,Li&al,Aref&al}, at least at not too low temperatures.

We begin by establishing our convention for assembling electron operators 
into Dirac spinors.
Assuming that superconductivity is due to a correlation between oppositely moving 
electrons in the same transverse mode, we expect it to show in correlation functions of
the operator
$\sum_{A=1}^N a_R^A b_L^A$, where the subscripts $R$ and $L$ designate the 
electron operators with positive and negative momenta, respectively. For this
to be a color singlet, $a_R$ should transform as $N$ of the $SU(N)$ and $b_L$ as
$\bar{N}$ (or vice versa). Thus, we define
a 2-component Dirac spinor $\psi$ as follows (omitting the color index $A$):
\be
\psi = \left[ \begin{array}{c} \sum_{k>0} (a_{Rk} e^{ikx} + b_{Rk}^\dagger e^{-ikx} ) \\
\sum_{k<0} (a_{Lk} e^{ikx} + b_{Lk}^\dagger e^{-ikx} ) \end{array} \right] \, .
\label{psi}
\ee
This is in the representation where the Dirac $\gamma$ matrices are given by
$\gamma^0 = \sigma_1$, $\gamma^1 = - i \sigma_2$, and $\gamma^5 = \sigma_3$, in terms
of the Pauli matrices {\boldmath $\sigma$}. The identification of the superconducting
channel as $a_R b_L$ implies
that $a_{Rk}^\dagger$ creates a $k>0$ electron, and $b_{Lk}^\dagger$ a
$k<0$ electron (i.e., $a_{Lk}^\dagger$ creates a $k<0$ hole). With this convention,
the upper and lower components of $\psi$ have opposite electric charges, 
the superconducting channel is $\bar\psi \psi$ (where $\bar{\psi} = \psi^\dagger \gamma^0$),
and an external electromagnetic potential couples to the axial current
$\bar{\psi} \gamma^\mu \gamma^5 \psi$. 

To reproduce the physics of such a superconductor on the gravity side, we consider 
the system of $N$ coincident D3 branes and a single D5 brane intersecting over a line.
As common in applications of the gauge/gravity duality, in the large $N$, large 't Hooft
coupling limit the
D3 branes are replaced by their classical geometry while the D5
is considered as a probe, i.e., its effect on the geometry is neglected.
The resulting geometry is that of a throat, of coordinate length $R$, pulled by 
the D3s out of the flat 10-dimensional (10d) spacetime:
\[
ds^2 = \frac{1}{\sf} \left( - dt^2 + d {\bf x}^2 \right) +
\sf \left( d \D^2 + \D^2 d \phi^2 \right) +
\]
\be
{} + \sf \left( d\rho^2 + \rho^2 d\Omega_3^2 \right) \, ,
\label{metric}
\ee
where 
\be 
f = 1 + \frac{R^4}{(\D^2 + \rho^2)^2} \, ,
\label{f}
\ee
$t$ and ${\bf x}=(x^1, x^2, x^3)$ are coordinates 
on the D3 worldvolume, $\D$ and $\phi$ are polar coordinates in the $(x^8, x^9)$ plane,
$\rho^2= (x^4)^2 +\ldots + (x^7)^2$, and $d\Omega_3^2$ is metric on a unit 3-sphere. 
In what follows, we use dimensionless units in which $R=1$.
The D3s are located at $\D=\rho=0$, where the metric (\ref{metric})
has a degenerate horizon. This metric is suitable for calculations
at zero temperature, the only case for which we present 
detailed  calculations here.

Matter in
the fundamental representation of $SU(N)$ (in our case, the electrons) 
is described by strings stretching between the D3s and a probe brane \cite{Karch:2002sh}.
We consider the case when the probe D5 wraps $x^1$, $\rho$ and the 3-sphere. 
Since $x \equiv x^1$ is the only spatial 
direction shared by the D5 and D3s, 
this brane intersection describes a theory of electrons
that live on a (1+1)-dimensional defect but interact via a (3+1)-dimensional non-abelian
gauge field. Overall the setup is similar to the system
of D3 and D7 branes intersecting over a plane, in which the electrons are confined
to move in {\em two} spatial dimensions \cite{Rey:2008zz,Davis:2008nv}. Note
that in our case there are two directions, 
$x^8$and $x^9$, orthogonal 
to all branes. In complex notation, the displacement of the
D5 relative to the D3s in the $(x^8,x^9)$ plane is $\D e^{i\phi}$ and forms an order
parameter suitable for description of superconductivity. The minimal distance between 
the D5 and D3s is the quasiparticle gap (in string units).

For a general D5 embedding, 
\ba
\D & = & \D(t, x, \rho, \alpha_i) \, , \label{D} \\
\phi & = & \phi(t, x, \rho, \alpha_i) \, , \label{phi}
\ea
where $\alpha_i$ are angles on the 3-sphere. Vibrations of the brane correspond to 
collective
modes of the electron fluid. In the 't Hooft limit ($N\to \infty$ at fixed 
$\lambda =g_s N$, where $g_s$ is the closed string coupling),
quantum fluctuations of the brane are suppressed by the large value of the brane tension,
so to the leading order the brane can be considered as a classical object. 
One can then explore various embedding ansatzes, which will typically have less coordinate
dependence than the most general form (\ref{D})--(\ref{phi}). We assume throughout that
$x^2 = x^3 =0$. In addition, all embeddings
we consider here are independent of $\alpha_i$. 
With this restriction, the Dirac-Born-Infeld (DBI) action of the D5 brane is
\be
S_{\rm DBI} = - 2\pi^2 T_5 \int dt  dx d\rho \rho^3 f^{3/4} [-\det(G_{ab} + F_{ab})]^{1/2} \, ,
\label{SD5}
\ee
where $T_5$ it the brane tension, $G_{ab}$ for $a,b=t,x,\rho$ are the components of the
induced metric, and $F_{ab} = \partial_a A_b - \partial_b A_a$
is a $U(1)$ gauge field (distinct from the usual electromagnetic field) on the 
D5 worldvolume. 
The classical dynamics of the brane is described by the Euler-Lagrange (EL)
equations following from the total of (\ref{SD5}) and a Wess-Zumino term that describes
the coupling of the D5 to the background Ramond-Ramond field \cite{Craps:1999nc}.

A spatially uniform winding of the order parameter can be described by (static) 
embeddings for which $\phi(x) = q x$, where $q$ is a constant, 
while $\D$ depends on $\rho$ only.
The constant $\partial_x \phi$ sources worldvolume electric field $F_{t\rho}$ through
the Wess-Zumino coupling. This leads to
a system of coupled equations for $\Delta$ and $F_{t\rho}$. We do not consider this 
case further here and move on to description of states in which there is initially no winding,
i.e., all current is carried by the normal component.

We begin with static $x$-independent embeddings $\D = \D(\rho)$, $\phi=0$, 
$A_t=A_t(\rho)$ with all other components of $A_a$ equal to zero. 
As explained in \cite{Kobayashi:2006sb} (in the context of a different D-brane system),
the value $\mu = A_t(\infty)$ is the chemical potential for
the density $\psi^\dagger \psi$. Turning to (\ref{psi}), we see that 
the $k>0$ and $k<0$ electrons are oppositely charged with respect to $\mu$. Thus,
in the superconductor, $\mu$ is conjugate to the electric current.
The lines of the radial field
$F_{t\rho}$ have nowhere on the brane to end, so the D5 must extend behind the horizon 
\cite{Kobayashi:2006sb}.
Hence the boundary condition $\D(\rho = 0) = 0$:
at any nonzero current the superconductor is gapless. 

The action (\ref{SD5}) for this type of embedding is $S_{\rm DBI} = - \int dt \cF$, 
where $\cF$ is the free energy:
\be
\cF = 2\pi^2 T_5 \int dx d\rho \rho^3 \sqrt{f} (1 + \D_{,\rho}^2 - F_{t\rho}^2)^{1/2} 
\, .
\label{cF2}
\ee
We use notation $\D_{,\rho}\equiv \partial_\rho \D$. The Wess-Zumino term vanishes.
The equation of motion for $A_t$ shows that the current 
\be
J = \frac{1}{2\pi^2 T_5} \frac{\delta \cF}{\delta F_{t\rho}} 
\label{I}
\ee
is $\rho$-independent.
Following the usual procedure of Legendre transforming to go from 
fixed $\mu$ to fixed $J$, we obtain the Legendre transformed free energy
\be
\widehat{\cF} = 
2\pi^2 T_5 \int dx d\rho (\rho^6 f + J^2)^{1/2} (1 + \D_{,\rho}^2)^{1/2} \, .
\label{legendre}
\ee
For each value of $J$, solutions to the corresponding
EL equation form a one-parametric family with $\D_{,\rho}(0)$ as 
a parameter. The general behavior
of a solution at large $\rho$ is a constant: $\D(\infty) = b$. 

\begin{figure}
\begin{center}
\includegraphics[width=3.25in]{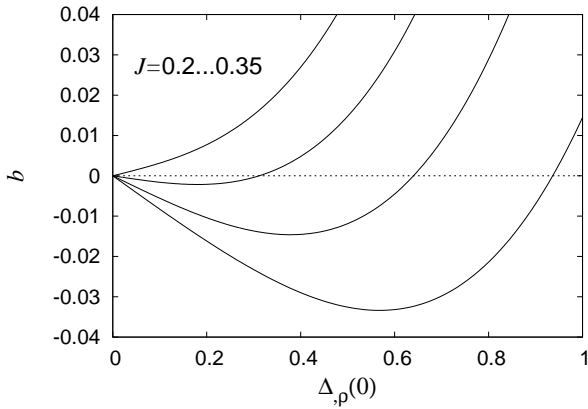}
\end{center}                                              
\caption{Equation-of-state curves for several values of the current $J$. $b$ is the 
asymptotic value of the D5 embedding, and $\D_{,\rho}(0)$ is its slope at the horizon.
The current increases in increments of 0.05 from bottom to top.
}                                              
\label{fig:eos}                                                                       
\end{figure}

We refer to the
dependence of $b$ on $\D_{,\rho}(0)$ as an equation-of-state (EOS) curve. These curves
provide a convenient way to study the phase transition. Some representative ones
are shown in Fig.~\ref{fig:eos}. 
A nonzero asymptotic value $b\neq 0$ corresponds to an explicit breaking of the
symmetry that shifts $\phi$ by a constant.
Since we are interested in solutions that break this
symmetry spontaneously, rather than explicitly, we look for nontrivial
solutions with $b=0$.
In Fig.~~\ref{fig:eos}, such a solution corresponds to a second zero (if any) of
the EOS curve; the first zero, at $\D_{,\rho}(0) = 0$,
is the trivial solution. We see that second zeroes exist at smaller currents but as
$J$ increases they decrease in magnitude and eventually merge with the
trivial solution and disappear. Numerically, the value of $J$ at which that happens
is $J_c = 0.3197$.

The way the nontrivial solution disappears at $J=J_c$ is reminiscent of a continuous
phase transition. In our case, however, it is only the $x$-independent superconducting
solution that disappears, while some $x$-dependent ones must remain. To see that,
consider the
linearized EL equations with time and space dependence; these can be obtained by Legendre
transforming the quadratic part of the D5 action. Upon substitution
$\Psi(t,x,\rho) = e^{-i\omega t + ikx} \D(\rho)$ for the position of the D5 in the $(x^8,x^9)$
plane, the linearized equation for $\D$ reads
\be
\frac{1}{\sqrt{C}} \partial_\rho \left( \sqrt{C} \D_{,\rho} \right) +
\omega^2 f_0 \D - \frac{k^2 \rho^6 f_0^2}{C} \D + \frac{2\D}{C} 
+ \frac{4 k J}{\rho^2 C} \D = 0 \, ,
\label{lin}
\ee
where $C= \rho^6 f_0 + J^2$ and $f_0 = 1 + 1/\rho^4$; the last term on the left-hand side
is from the WZ action.
At $\rho\to\infty$, this reduces to the spherical wave equation in (3+1) dimensions.
At $\rho\to 0$, and
$\omega \neq 0$, the leading asymptotics are $\D(\rho)\sim \rho e^{\pm i \omega / \rho}$.
We choose the positive sign, corresponding to waves falling into the horizon.

We consider real $k$ and complex $\omega$.
Unstable modes of the trivial $\D\equiv 0$ embedding are eigenmodes of 
(\ref{lin}) with $\iome > 0$. They decay exponentially
at both $\rho\to 0$ and $\rho \to \infty$. For these boundary conditions, 
$\omega^2$ is purely real; 
hence $\omega$ is purely imaginary. Let us return for a moment to the spatially
uniform case $k = 0$. By numerically solving (\ref{lin}) with Dirichlet boundary 
conditions, we find that, when $J < J_c$ but close to it, there
is only one unstable mode, and its profile closely matches that of the nontrivial 
static solution to the nonlinear problem (\ref{legendre}). This is evidence that
the instability of the trivial embedding develops into one of the nontrivial embeddings
we considered earlier. At $J=J_c$, the frequency crosses zero in a smooth, analytic
manner, approximately as $\iome = -0.9 (J - J_c)$, and the instability disappears.

On the other hand, for $k \neq 0$ and $J > 0$, the change of variable to
$z = 1/\rho$ brings the small $\rho$ (large $z$) limit of
(\ref{lin}) to the form
\be
- \frac{1}{z^2} \partial_z (z^2 \D_{,z}) - \frac{\gamma}{z^2} \D
= \omega^2 \D \, ,
\label{asymp}
\ee
where $\gamma = 4 k / J - (k/J)^2$. This is a Schr\"{o}dinger equation with 
a fall-to-center 
potential. The potential is supercritical (i.e., the full eq.~(\ref{lin})
has an infinite number of bound 
states) when $\gamma > 1/4$ \cite{Landau&Lifshitz}. There is always a band
of $k$ satisfying this condition. We conclude that the normal state remains unstable
even at $J > J_c$, but the instability now occurs only in channels with nonzero 
winding. We interpret this as an indication that in a stable state at $J > J_c$ at least
part of the total current is a supercurrent.

For calculations at a finite temperature, $T$, the region near $\Delta = \rho = 0$ 
is replaced by a black hole \cite{Maldacena:1997re}. This cuts off the large
$z$ region in (\ref{asymp}), which for fall-to-center problems typically reduces 
both the number and the growth
rates of the unstable modes. It is of interest to
consider the possibility that, at sufficiently high $T$, there will be a depairing
transition associated with a near-critical behavior in the vicinity of some $J_c(T)$.
Indeed,
experimentally, superconducting nanowires have been observed to switch to the
normal state at currents below
the estimated depairing current \cite{Sahu&al,Li&al,Aref&al}. 
Large fluctuations of the switching current found in these experiments have
been interpreted \cite{Sahu&al,Li&al,Aref&al} as a
consequence of phase slips. This interpretation is consistent with a second-order (or
nearly so)
transition at some $J= J_c(T)$, 
as it implies that the free energy barrier suppressing 
phase slips can be almost completely removed by bringing the current close to the 
estimated depairing current.

The following estimate suggests that, if the observed fluctuations of the switching 
current are indeed a result of a near-critical behavior at $J \approx J_c(T)$,
that behavior is
controlled by a Gaussian (not necessarily stable) fixed point, at least
at temperatures where thermally activated phase slips are thought to be the main effect.
The exponential factor in the rate of thermal activation is 
$\exp(-\delta \cF / T)$, where $\delta \cF$ is the free energy barrier and $T$ is the 
temperature. Near $J=J_c$, $\delta \cF$ scales as the product of the free energy 
density and the correlation length, i.e., as $(J_c - J)^{2 - \alpha - \nu}$, in terms of
the conventionally defined critical exponents. For the Gaussian point,
$\alpha = 0$ and $\nu = \half$, so $\delta \cF \sim (J_c-J)^{3/2}$. Curiously, this
is the same scaling as obtained for a Josephson junction \cite{Fulton&al}.
It has been found to provide a good fit to the data in Ref.~\cite{Li&al} and for the
amorphous samples in Ref.~\cite{Aref&al} (for the crystalline samples \cite{Aref&al}, 
the $5/4$ power law has been found to be a better fit).
In contrast, for a fixed point obeying hyperscaling, 
$2 - \alpha - \nu = 0$ and $\delta \cF$ scales to a constant.

The author thanks A.~Bezryadin for a discussion. 
This work was supported in part by the U.S. Department of Energy
through grant DE-FG02-91ER40681.

\end{document}